\magnification=\magstep1
\baselineskip=17pt
\font\smallrm = cmr8 
\hfuzz=6pt

$ $

\vskip 1cm

\centerline{\bf Quantum polar decomposition algorithm}

\bigskip

\centerline{Seth Lloyd,$^{1,2*}$ Samuel Bosch,$^3$
Giacomo De Palma,$^{1,2}$ Bobak Kiani,$^1$ 
Zi-Wen Liu,$^4$}
\centerline{ Milad Marvian,$^{1,2}$ Patrick Rebentrost,$^5$ 
David M. Arvidsson-Shukur$^{6,2}$ }
\bigskip
{\smallrm
\centerline{1. Department of Mechanical Engineering, MIT,  
2. Research Lab for Electronics, MIT,}
\centerline{ 3. Narang Lab, Engineering And Applied Sciences, 
Harvard University, }
\centerline{ 4. Perimeter Institute, 5. Center for Quantum
Technologies, National University of Singapore,}
\centerline{ 6. Cavendish Laboratory, Department of Physics, 
University of Cambridge }
\centerline{$^*$ to whom correspondence should be addressed: slloyd@mit.edu}
}

\vskip 1cm

\noindent{\it Abstract:} The polar decomposition for a matrix
$A$ is $A=UB$, where $B$ is a positive Hermitian matrix and $U$ is 
unitary (or, if $A$ is not square, an isometry).  This paper shows that the
ability to apply a Hamiltonian $\pmatrix{ 0 & A^\dagger \cr A & 0 \cr} $ 
translates into the ability to perform the transformations $e^{-iBt}$ and
$U$ in a deterministic fashion.  We show how to use the quantum polar
decomposition algorithm to solve the quantum Procrustes problem, to perform 
pretty good measurements,
to find the positive Hamiltonian closest to any Hamiltonian, and
to perform a Hamiltonian version of the quantum singular value transformation.

\vskip 1cm

The polar decomposition of a matrix is $A = UB = \tilde B  U$,
where $B = (A^\dagger A)^{1/2}$, $\tilde B =  (A A^\dagger)^{1/2}$ are 
positive Hermitian matrices, and $U = A (A^\dagger A)^{-1/2}
 =  (A A^\dagger)^{-1/2} A$ is unitary (when $A$ is
square) or an isometry (when $A$ is not square).  
The polar decomposition has many applications in linear algebra [1-2].   
$U$ is in essence
the `closest' unitary to $A$: it is the unitary that minimizes
$\| U - A\|_2$ in Frobenius norm.   In a quantum mechanical setting,
it would be useful to have a method which 
allows one to perform the polar decomposition.   This paper generalizes
the quantum linear systems algorithm [3-4] to construct
just such a quantum algorithm for the polar decomposition.   In
particular, given the ability to apply a Hamiltonian $H =
\pmatrix{ 0 & A^\dagger \cr A & 0 \cr} $,
we show how to apply the unitary/isometry $U$ in time $O(\kappa)$, 
where $\kappa$ is the condition number of $A$, and how to apply
the transformations $e^{-iBt}$, $e^{-\tilde Bt}$ in 
time $O(\kappa t/\epsilon)$, 
where $\epsilon$ is the accuracy
to which these transformations are to be performed.   The time taken
to perform the quantum polar decomposition is independent of the dimension
of the Hilbert space on which $H$ acts: for example $A$ could 
be an operator acting on the infinite dimensional Hilbert space for a
collection of modes of the electromagnetic field.  

We apply the algorithm 
to reconstruct unitary transformations from examples of input-output pairs
-- the quantum Procrustes problem [2], 
to perform pretty good measurements, and to find the positive matrix
closest to any Hermitian matrix [1].  In its generalized form, 
the quantum polar decomposition
algorithm can be considered to be a Hamiltonian version of the
quantum singular value decomposition [5].

\bigskip\noindent{\it  Preliminaries}

Write the $m\times n$ matrix $A$ as
$A = \sum_j \sigma_j \ell_j r_j^\dagger$, 
where $\sigma_j$
are the singular values of $A$ and $\ell_j \in {\cal H}_L =C^m $, $r_j
\in {\cal H}_R = C^n $ are the corresponding left
and right singular vectors.   
The polar decomposition matrices
are then $U= \sum_j \ell_j r_j^\dagger$, 
an isometry from ${\cal H}_R$ to ${\cal H}_L $,
$B = \sum_j \sigma_j r_j r_j^\dagger$, acting on ${\cal H}_R$, and 
$\tilde B = \sum_j \sigma_j \ell_j \ell_j^\dagger$, acting on ${\cal H}_L $.   
Suppose that one has
the ability to apply a Hamiltonian $\pm H$, where $H = A + A^\dagger$
acts on the Hilbert space ${\cal H}_R \oplus {\cal H}_L$.
$H = A+ A^\dagger$ can also be represented in matrix form as 
$$H = 
\pmatrix{ 0 & A^\dagger \cr A & 0 \cr}.  \eqno(1) $$    
We show how to use
the ability to apply the Hamiltonian $\pm H$ to perform the transformations
 $e^{-iBt}$, $e^{-i\tilde Bt}$ and
$U$, in a deterministic fashion.  
While we will present the quantum polar transformation in terms of
matrices, we also show that the entire construction of the
quantum polar decomposition goes through when $A$ is an operator
on an infinite dimensional space, e.g., a polynomial in annihilation
and creation operators on a set of harmonic oscillators.         

\bigskip\noindent{\it Generalized quantum linear systems algorithm}

The quantum polar decomposition algorithm is based on a deterministic
extension of the original non-deterministic quantum 
linear systems algorithm [3].
The generalized version of the original quantum linear 
systems algorithm operates as follows.  
For a generic Hermitian $H$ with eigenvectors $|j\rangle$ and eigenvalues
$\lambda_j$, we first review how to perform the
transformation 
$$|\psi\rangle \rightarrow  e^{-i f(H)} |\psi\rangle
= \sum_j e^{-i f(\lambda_j)} \psi(j) |j\rangle,\eqno(2)$$ 
where $|\psi\rangle = \sum_j \psi(j) |j\rangle$.
Here $f$ can be any computable function of the eigenvalues. 
To accomplish the transformation in equation (2), we employ 
the quantum phase estimation algorithm [6] --
a digitized version of the von Neumann pointer variable
model of measurement [7] --  to correlate the 
eigenvectors of $H$ with estimates of the corresponding eigenvalues:
$$|\psi\rangle \otimes| x = 0\rangle \rightarrow
\sum_j \psi(j) |j\rangle \otimes |\tilde \lambda_j\rangle,\eqno(3)$$
where $\tilde \lambda_j $ is a $b$-bit approximation to 
$\lambda_j$.  
Now multiply each term by a phase
that is a function of the estimated eigenvalue to obtain
$$\sum_j\psi(j) e^{-i   f(\tilde \lambda_j)} 
|j\rangle \otimes |\tilde \lambda_j\rangle.\eqno(4)$$
Undoing the quantum phase estimation algorithm yields the desired
transformation of equation (2) to $b$ bits of accuracy in $\lambda_j$.

To perform the transformation (2) in infinite dimensional systems,
adjoin a continuous variable with position/momentum operators $X,P$:
$[X,P]=i$.   Perform von Neumann's Hamiltonian pointer variable model 
of measurement [7]: apply the Hamiltonian $H\otimes P$ to
the initial state $|\psi\rangle\otimes |x=0\rangle$ for 
a unit time interval $t=1$ to create the state
$\sum_j \psi_j |j\rangle \otimes |\lambda_j\rangle$, the
continuous variable version of equation (3).   
Now use methods of continuous variable quantum computation [8]
to apply the Hamiltonian $f(X)$ to the pointer variable, 
again for unit time, yielding the state 
$\sum_j \psi_j e^{-if(\lambda_j)} |j\rangle \otimes |\lambda_j\rangle$.   
If $f$ is a $q$th order polynomial
in $X$ this step takes time $O(q)$.  
Finally, apply the Hamitonian $-H\otimes P$ for unit time. 
The result is the state
$$ e^{iH\otimes P} (I\otimes e^{-if(X)}) e^{-i H\otimes P}
|\psi\rangle\otimes | x=0\rangle 
= e^{-if(H)}|\psi\rangle\otimes |x=0\rangle, 
\eqno(5)$$ 
which is again the desired transformation of equation (2), but
now in the context of infinite-dimensional operators rather
than finite-dimensional matrices.

\vfill\eject
\bigskip\noindent{\it Quantum polar transformation algorithm}

Framed in the background of the generalized linear systems algorithm,
the quantum polar transformation immediately reveals itself. 
To perform the polar transformation unitary/isometry $U$,
take $f(x) = ( \pi/2) 
(1- {\rm sign}(x))$ in equation (2).  This transformation multiplies an
eigenvector of $H$ by the sign of its eigenvalue.
The eigenvectors of $H$ take the form
$$\pmatrix{ 0 & A^\dagger \cr A & 0 \cr} \pmatrix{ r_j\cr 
\pm \ell_j} = \pm \sigma_j \pmatrix{ r_j\cr 
\pm \ell_j}.\eqno(6)$$
Using the relationships given above between the eigenvectors and 
eigenvalues of $H$ with the left and right singular vectors and the singular
values of $A$, we see that
$$ \pmatrix{ r_j \cr 0} = (1/2) \bigg( \pmatrix{ r_j \cr \ell_j} 
+ \pmatrix{ r_j \cr - \ell_j} \bigg)
\longrightarrow
(1/2) \bigg( \pmatrix{ r_j \cr \ell_j}
- \pmatrix{ r_j \cr - \ell_j} \bigg) = \pmatrix{ 0\cr \ell_j}.\eqno(7)$$
Similarly, 
$$\pmatrix{ 0 \cr \ell_j} \longrightarrow \pmatrix{ r_j \cr 0}.\eqno(8)$$
Writing the outcome of the transformation in both matrix
and quantum bra-ket notation,
we see that the outcome of the generalized quantum linear systems algorithm 
is the state
$$ \big(\sum_j |\ell_j\rangle \langle r_j| +
\sum_j |r_j\rangle \langle \ell_j| \big) |\psi\rangle = 
(U + U^\dagger)|\psi\rangle = \pmatrix{ 0 & U^\dagger \cr U & 0} 
\pmatrix{ |\psi\rangle_R \cr |\psi\rangle_L }.
\eqno(9)$$
Thus, as promised, we can perform the unitary/isometry $U$ 
on ${\cal H}_R$, and $U^\dagger$ on ${\cal H}_L$, where $U$ is
the unitary/isometry in the polar decomposition of $A$.

Without loss of generality, take the largest singular value of $A$ to be one.
The condition number of $A$ is then the inverse of the smallest singular
value $\kappa = \sigma_{min}^{-1}$.    To perform the quantum phase
estimation algorithm sufficiently accurately to resolve the sign
of the smallest eigenvalue then takes time $O(\kappa)$.

Following reference [3], we can also perform $U$ only on the
well-conditioned subspace of $A$.    When the quantum phase algorithm
yields an estimate of the singular value that is
smaller than some value $1/\tilde\kappa$, we decline to change its sign,
and flip a `flag' qubit, initially in the state $|0\rangle$,
to the value $|1\rangle$.    This procedure
allows us to project onto the well-conditioned subspace spanned
by singular vectors whose singular value is greater than 
or equal to the inverse of the
chosen effective condition number $\tilde\kappa$, and to perform $U$ only on
this subspace.   The operation performs 
$$U_{\tilde\kappa} + U_{\tilde\kappa}^\dagger 
= \sum_{j: \sigma_j \geq 1/\tilde\kappa} |\ell_j\rangle \langle r_j| 
+  |r_j\rangle \langle \ell_j|\eqno(10)$$
on the well-conditioned
 subspace, and the identity on the poorly conditioned subspace,
raising a flag if the operation has projected the initial state
on the poorly conditioned subspace.  The algorithm takes
time $O(\tilde \kappa)$. 

Now take $f(x) = |x|t$ in the quantum phase estimation algorithm.
The resulting transformation is 
$$ \big( e^{-i \sqrt{ A^\dagger A} t} + e^{-i \sqrt{A A^\dagger} t} \big)
 |\psi\rangle. \eqno(11)$$
To perform this transformation to accuracy $\epsilon$ takes time
$ O(\kappa t/\epsilon)$.  So we can perform the transformations
$e^{-iBt}$ on ${\cal H}_R$ and $e^{-i\tilde Bt}$ on ${\cal H}_L$ as well.
QED $=$ {\it Quod Erat Demonstrandum} 
 $=$ Quite Easily Done.   

\bigskip\noindent{\it Generalization}

Writing $f(A) = \sum_j f(\sigma_j) |\ell_j \rangle \langle r_j|$, the same
methods allow us to apply any Hamiltonian of the form
$$ f(A) + f(A)^\dagger, \quad f( \sqrt{A^\dagger A}) + f(\sqrt{A A^\dagger}) 
. \eqno(12)$$ 
In addition, we can apply any Hamiltonian of the form
$f(K)+ f(K^\dagger)$, where $K, K^\dagger$ have the same eigenvectors as
$\sqrt{A^\dagger A}$, $\sqrt{A A^\dagger}$, and the same eigenvalues up
to a sign $\pm 1$.

\bigskip\noindent{\it Applications}

The polar decomposition is widely applicable for problems where we
wish to find the closest matrix of a particular form to a given matrix [1]. 
When $A$ is Hermitian, then
$B=\tilde B$ is the closest positive Hermitian matrix to $A$ in
Frobenius norm.  Similarly, $U$ is the
solution to the problem of minimizing $\|U - A\|_2$ 
over all
unitaries/isometries.   We now apply the quantum polar decomposition to the
problem of finding and applying a unitary transformation given 
examples of input/output pairs, and to performing pretty good 
measurements.    We show that the polar decomposition
algorithm can be thought of as a Hamiltonian version
of the quantum singular value transformation.

\bigskip\noindent{\it Recreating a unitary from input/output pairs}

Suppose that we are given $r$ input/output pairs
$$\{ ( |\phi_j\rangle, |\psi_j\rangle ) \} \in {\cal H}_0 \oplus
{\cal H}_1. \eqno(13)$$ 
Assume that we can coherently apply the transformations
$$ |j\rangle\otimes|0\rangle \rightarrow |j\rangle\otimes |\phi_j\rangle,
\quad |j\rangle\otimes|0\rangle \rightarrow |j\rangle\otimes |\psi_j\rangle,
\eqno(14)$$
where $|j\rangle$ are computational basis states. 
We  wish to find
the unitary/isometry $U$ that minimizes
$$\| U F - G \|_2^2,\eqno(15)$$  
where $F$ is the matrix whose columns are the $\{ |\phi_j\rangle \}$,
and $G$ is the matrix whose columns are the $\{ |\psi_j\rangle \}$.
This is the quantum Procrustes problem [2].   (Procrustes was a mythical
bandit in ancient Greece, who would compare anyone he captured
to the size of his bed, and then either stretch out or chop off the
parts he needed to make the captive fit.   Here, we are trying
to fit $F$ as non-violently 
as possible to $G$ by a unitary/orthogonal transformation.)

To minimize the distance of $U$ from the desired transformation, equation (15),
subject to the constraint that $U$ is a unitary/isometry, define a Lagrangian
$${\cal L} = \sum_{j=1}^r \| |\psi_j\rangle - U|\phi_j\rangle \|^2
- {\rm tr} \Lambda(U U^\dagger -I), \eqno(16)$$
where $\Lambda$ is a positive Hermitian matrix of Lagrange
multipliers.    Taking the variation of ${\cal L}$ with respect
to $U$ and to $\Lambda$, we find that the extremum is attained
for $U = A(A^\dagger A)^{-1/2}$, where 
$A = \sum_{j=1}^r |\psi_j\rangle \langle \phi_j|$.   That is, the solution
to the Procrustes problem is the unitary $U$ 
of the polar decomposition of the matrix $A$
relating inputs to outputs.

Note that unitaries of the form $U = A \tilde K$, 
where $\tilde K$ has the same eigenvectors 
as $(A^\dagger A)^{-1/2}$ and the same eigenvalues up to a sign $\pm 1$, 
are also solutions to the Lagrange equations.  Direct substitution
into equation (15) shows that $\tilde K = (A^\dagger A)^{-1/2}$ gives
the unitary/isometry with minimum distance to $A$,  
with the other solutions corresponding to local maxima.  The global
maximum (reverse Procrustes problem [2]) occurs when
all the eigenvalues are negative: $-U$ maximizes the distance in equation
 (15).

To apply the Hamiltonian $ H = A+A^\dagger$ we use our quantum
access to the input output pairs to create the state
$${1\over\sqrt{2r}}\sum_{j=1}^r |j\rangle \otimes  
\pmatrix{ |\phi_j\rangle \cr  |\psi_j\rangle }.
\eqno(17)$$
Tracing out the first register, we find the second 
register is in the state given by the density matrix 
$$ \rho_A = {1 \over 2r}\pmatrix{ C&A^\dagger\cr A & \tilde C}, \eqno(18)$$
where $C =(1/2r) \sum_j |\phi_j\rangle \langle \phi_j|$ and
$\tilde C = (1/2r) \sum_j |\psi_j\rangle \langle \psi_j|$.
Let $P_0$,$P_1$ be the projectors onto ${\cal H}_0$,
${\cal H}_1$ respectively, 
By applying the unitary transformation $V = P_0 - P_1$, 
we can also create the state,
$$ \tilde \rho_A =  V \rho_A V^\dagger 
= {1\over 2r}\pmatrix{ C& -A^\dagger \cr - A  & \tilde C}. \eqno(19)$$
Now we use density matrix exponentiation [9] to apply the 
infinitesimal transformation
$$\eqalign{ e^{i\Delta t \tilde \rho_A } e^{-i\Delta t \rho_A}
&= I +i\Delta t \tilde\rho_A -i\Delta t \rho_A  + O(\Delta t^2) \cr
  &=  I -i\Delta t(A+A^\dagger)/r   + O(\Delta t^2) 
= e^{-i\Delta t(A+A^\dagger)/r} + O(\Delta t^2).\cr}
\eqno(20)$$ 
That is,
we can effectively apply the Hamiltonian $H = A + A^\dagger$, 
and so can apply the quantum Procrustes transformation 
$U$ on any state $|\chi\rangle$ via the quantum
polar decomposition algorithm. 
The algorithm takes time $O(r\kappa)$, where $r$ is
the number of input-output pairs and
$\kappa$ is the condition number of 
$A = \sum_{j=1}^r |\psi_j\rangle \langle\phi_j|$.
An even more general solution to the quantum Procrustes problem is
provided by the quantum emulation algorithm [10], which does not require
the assumption that 
we can access the input-output state pairs in quantum superposition.

\bigskip\noindent{\it Hamiltonian quantum singular value transformation
}

The method used in the previous section shows how to perform a Hamiltonian
version of the quantum singular value transformation [5].   Let $A$
be any $m\times n$  off-diagonal block of a Hamiltonian that we are able
to apply.  As we are only interested in the time evolution on the subspace
acted on by $A$ and $A^\dagger$, without loss of generality
we can simply consider Hamiltonians acting on ${\cal H}_0 \oplus {\cal H}_1$ 
of the form $\pm M$, where
$$  M =  \pmatrix{
D&   A^\dagger\cr
A &  \tilde D},\eqno(21)$$ 
where $D,\tilde D$ are arbitrary Hermitian matrices.
Now use the same trick as above: 
apply $ M$ for time $\Delta t$,
followed by the unitary transformation
$V=P_0 - P_1$, where $P_0,$ $P_1$ are the projectors onto ${\cal H}_0,$
 ${\cal H}_1$. 
The resulting transformation is equivalent to the application
of a Hamiltonian   
$$  V M V^\dagger =  \pmatrix{ D&  -  A^\dagger\cr
- A &  \tilde D}.\eqno(22)$$  
By the first order Trotterization trick of equation (19) above, and
noting that we can apply $\pm M$ and so also $\pm VMV^\dagger$, we
can then apply the effective Hamiltonian
$$  (M - VMV^\dagger )/2  =   \pmatrix{ 0&    A^\dagger\cr
 A &  0}.\eqno(23)$$
The ability to apply this Hamiltonian then translates into the ability
to apply any Hamiltonian with the same singular vectors as $A$, and whose
singular values are some computable function of the singular values
of $A$ as in equation (12) above.  The Hamiltonian singular value
transformation can be extended to infinite dimensional systems
via the techniques of equation (5) above.    It is an open
question whether it is possible to use the methods of references [4-5] to 
forego the use of the quantum phase estimation algorithm and
to apply such Hamiltonians more efficiently.

\bigskip\noindent{\it Pretty good measurements}

Suppose that we wish to enact the pretty good (square root) measurement
to distinguish between $n$ states $\{ |\phi_j\rangle \}$.  The pretty
good measurement for distinguishing between pure states consists of
projections onto the states 
$$ |\chi_j\rangle = 
( \sum_j |\phi_j\rangle \langle \phi_j|)^{-1/2} |\phi_j\rangle.\eqno(24)$$   
That is, as shown in [11], 
the pretty good measurement is simply a von Neumann measurement that
projects onto the rows of $U = A (A^\dagger A)^{-1/2} $, where
$A = \sum_j |j\rangle \langle \phi_j|$ is the matrix whose rows are 
$\{ \langle \phi_j| \}$. 

This measurement can be performed efficiently using the quantum
polar decomposition algorithm.   In the algorithm from the previous
section for performing
the polar decomposition unitary $U$ given quantum access to the input
output pairs $\{ ( |\phi_j\rangle, |\psi_j\rangle ) \}$, 
take $|\psi_j\rangle = |j\rangle$.  To perform the pretty good
measurement on a state with density matrix $\rho$, first perform
the unitary $U$ then measure in the $|j\rangle$ basis.   The result
$|j\rangle$ occurs with probability
$$ p(j) = \langle j| U \rho U^\dagger |j\rangle =
\langle j| A(A^\dagger A)^{-1/2} \rho (A^\dagger A)^{-1/2} A^\dagger |
j \rangle = \langle \chi_j | \rho | \chi_j\rangle, 
\eqno(25)$$
which from equation (24) are just the probabilities for the pretty good
measurement.  To re-prepare the output state $|\chi_j\rangle$ -- that
is, to leave the system in the eigenstate of the pretty good measurement --
then simply apply the unitary $U^\dagger$ to the state $|j\rangle$.   
    
\bigskip\noindent{\it Conclusion:}  The polar decomposition of a matrix
$A$ is $A= U B = \tilde B U$, where $U = A(A^\dagger A)^{-1/2} $ is unitary, and
$B = (A^\dagger A)^{1/2}$, $\tilde B = (A A^\dagger)^{1/2}$ are
 positive Hermitian matrices.
 This paper showed how the ability to apply the Hamiltonian
 $\pmatrix{ 0 & A^\dagger \cr A & 0 \cr} $
translates into the ability to perform $e^{-iBt}$, $e^{-i\tilde Bt}$ and $U$ 
in a deterministic fashion.   The algorithm is based on the original
linear systems algorithm [3] and takes time $O(\kappa)$ to perform
$U$, where $\kappa$ is the condition number of $A$.   The time taken
to perform the algorithm is independent of the dimension of the
Hilbert space.   Indeed, the Hamiltonian $H$ could act on an infinite
dimensional Hilbert space, corresponding, e.g., to modes of the
electromagnetic field. 

The polar decomposition has many potential applications in 
linear algebraic protocols [1-2].  Here, we showed how to use the quantum
polar decomposition algorithm to perform
the optimal unitary that approximately reproduces a set of input/output
pairs (the quantum Procrustes problem), to apply the positive Hamiltonian
that is closest to a given Hamiltonian, and to perform pretty 
good measurements corresponding to a set of pure states.  
The method employed can be thought of as a Hamiltonian
version of the quantum singular value transformation [5].

\vskip 1cm
\noindent{\it Acknowledgments:} 
The authors thank Yann LeCun and Christian Weedbrook for helpful conversations.
This work was funded by AFOSR, ARO, DOE, and IARPA. 
D.R.M.A-S. received funding from the EPSRC and Lars Hierta’s Memorial Foundation.

\vskip 1cm
\noindent {\it References:}

\bigskip\noindent[1] N.J. Higham, {\it Siam J. Stat. Comput.} {\bf 7}, 1160-1174
(1986).

\bigskip\noindent[2] J.C. Gower, G.B. Dijksterhuis, {\it Procrustes Problems,}
Oxford University Press, Oxford (2004).

\bigskip\noindent [3] A.W. Harrow, A. Hassidim, S. Lloyd,
{\it Phys. Rev. Lett.} {\bf 103}, 150502 (2009).

\bigskip\noindent [4]  A.M. Childs, R. Kothari, R.D. Somma,
{\it SIAM J. Comput.} {\bf 46}(6), 1920–1950 (2017).

\bigskip\noindent [5] A. Gily\'en, Y. Su, G.H. Low, N. Wiebe,
{\it STOC 2019: Proceedings of the 51st Annual ACM SIGACT 
Symposium on Theory of Computing,} pages 193-204, June 2019.

\bigskip\noindent [6] A.Yu. Kitaev, `Quantum measurements and the 
Abelian Stabilizer Problem,' arXiv: quant-ph/951102

\bigskip\noindent [7] J. von Neumann, {\it Mathematische Grundlagen 
der Quantenmechanik,} Springer (1932); {\it 
Mathematical Foundations of Quantum Mechanics,} Princeton (1955).

\bigskip\noindent [8] S. Lloyd, S.L. Braunstein, 
{\it Phys. Rev. Lett.} {\bf 82}, 1784–1787 (1999).

\bigskip\noindent [9]  S. Lloyd, M. Mohseni, P. Rebentrost, 
{\it Nature Physics} {\bf 10}(9), 631-633 (2014); 
arXiv: 1307.0401.

\bigskip\noindent [10] I. Marvian, S. Lloyd,
`Universal Quantum Emulator,'
arXiv: 1606.02734.

\bigskip\noindent [11] V. Giovannetti, S. Lloyd, L. Maccone,
{\it Phys. Reva .A} {\bf 85}, 012302 (2012); arXiv: 1012.0386.

\vfill\eject\end